# Artificial Intelligence-Based Smart Grid Vulnerabilities and Potential Solutions for Fake-Normal Attacks: A Short Review

Jema David Ndibwile
Carnegie Mellon University | Africa
jeman@cmu.edu

*Abstract*

**Smart grid systems are critical to the power industry, however their sophisticated architectural design and operations expose them to a number of cybersecurity threats, such as data tampering, data eavesdropping, and Denial of Service, among others. Artificial Intelligence (AI)-based technologies are becoming increasingly popular for detecting cyber assaults in a variety of computer settings, and several efforts have been made to secure various systems. The present AI systems are being exposed and vanquished because of the recent emergence of sophisticated adversarial systems such as Generative Adversarial Networks (GAN). The purpose of this short review is to outline some of the initiatives to protect smart grid systems, their obstacles, and what might be a potential future AI research direction.**

## I. INTRODUCTION

A smart grid is an electrical system that incorporates a number of operations and energy measures, such as smart meters, smart appliances, renewable energy sources, and energy efficient resources [28]. Smart devices and the advancement of ICT technologies in general have influenced the transition from traditional grid to smart grid and, eventually, to state-of-the-art operations. However, security concerns like physical infrastructure destruction, data poisoning, denial of service, malware, and infiltration are introduced by complex architecture, which includes communication networks and multiple end devices [1]. As a result, there are a number of research efforts underway in academia and industry to solve these security concerns in order to secure the widespread deployment and survival of complex systems.

A review of broad assaults on smart grid systems, as well as machine learning algorithms to identify and mitigate these attacks, particularly false data injection (FDI) attempts, is described in this brief survey study. The challenges that these systems would have in identifying assaults like Generative Adversarial Networks (GAN), which can generate fake-normal traffic and so avoid detection [29], are also explored. The paper's key recommendation is that GAN attacks be explored further in order to figure out how to modify machine learning and other artificial intelligence technologies to detect them properly.

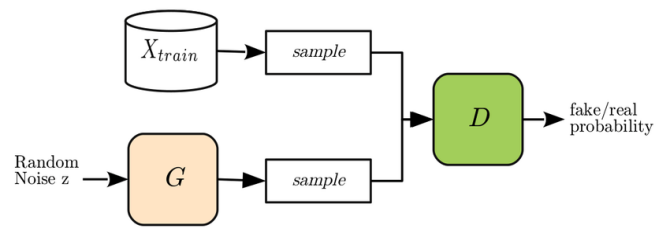

Fig. 1. Generative Adversarial Network [30]

## II. CLASSIFICATION OF SMART GRID ATTACKS

A number of studies have attempted to categorize smart grid assaults. To classify attacks, Rawat et al. [2] used types of networks that the attacks target, such as Home Area Network (HAN), Wide Area Network (WAN), and Neighborhood Area Network (NAN). Security objectives such as Confidentiality, Integrity, and Availability (CIA) are emphasized in these three categories.

Various authors, notably [3, 4], have examined trust, privacy, usability, and connectivity issues in great detail. According to the authors, security concerns have risen dramatically as a result of the widespread use and popularity of public networks. As a result, cybersecurity has become a critical component of the long-term viability of numerous systems. They go on to discuss the smart grid on public networks' commercial operations and system design. Based on their investigation, they provided numerous techniques for communications and field devices to address security needs, such as access control, secrecy, and integrity through various cryptographic algorithms. Wang et al. [5] classified various smart grid cyber assaults based on security objectives (CIA) and provided similar mechanisms via cryptographic algorithms, network protocols, and secure architecture.

Sun et al. [6] on the other hand, have thoroughly examined current smart grid systems such as SCADA and the security concerns they provide. They looked into prospective cybersecurity solutions, CPS cybersecurity testbeds, and unsolved cybersecurity difficulties. They also pointed out that present security methods are not always applicable to the smart grid context. They claim that the stringent availability standards of electricity systems often cause problems. In addition, the research looked into several Intrusion Detection System (IDS) methodologies that can be used in a variety of smart grid contexts and applications. Nonetheless, due to a lack of validation and verification on live operations and in the real world, most of these proposed methodologies and approaches have yet to be adopted by industry.

Adepu et al. [7] used the Electric Power and Intelligent Control (EPIC) testbed to examine the exploitation of vulnerabilities, designing and launching attacks utilizing several methodologies. According to the study, current attacks on transmission systems, loss of line and generators, and automatic generation controllers (AGC) are generally focused on limited locations [8, 9]. There has been minimal work put into addressing these vulnerabilities in dispersed systems or specific components that may be more vulnerable than highly secure transmission systems in general. For example, the likelihood of attack scenarios on regularly used renewable energy inverters using a standard home PC is one of the areas that has been overlooked.

Sagiroglu et al. [10] went on to categorize general vulnerabilities and threats in hardware, software, and networks related to Smart grid. The majority of the security vulnerabilities addressed in this research are those that are most frequent and applicable in everyday computing environments.

More specifically, Gunduz et al. [12] argue that the Advanced Smart Metering System in smart grids is more vulnerable to various cyber attacks than Smart Distribution Substations because Smart Meters are designed to exist on the utility network's edge, directly facing unprotected and uncontrollable HAN devices at the users' premises. The Distribution Substations, on the other hand, do not pose significant security risks despite their high level of automation and proximity to densely populated regions of the city since they are designed to be secure.

### III. FALSE DATA AND FAKE-NORMAL ATTACKS

There are a variety of security measures available to protect smart grid systems, having the following goals in mind:

- Smart grid components and protocols have inherent security flaws and vulnerabilities that must be addressed
- Reducing the negative impact of cyber-attacks and incursions on essential infrastructure
- Developing state-of-the-art deterrence tactics and periodically validating and improving them in order to identify new attacks

However, other assaults, such as false data injection and fake-normal traffic generation, are understudied and could pose a significant threat to a wide range of devices connected to the smart grid and its environment.

Cui et al. [11] offered a detailed assessment on the evolution of machine-learning in identifying bogus data injection attacks, with a focus on the attacks. Due to the obvious complex structure of smart grids, such as the deep integration of electrical components, false data injection attacks could be carried out effectively.

The transmission and distribution networks, power generation systems, and communication connections are all possible targets for a fake data injection attack. Furthermore, these attacks can be classified as replay/man-in-the-middle, zero dynamics, and FDI assaults.

The multiple end devices at various smart grid interfaces, according to Wei et al. [13], enable the smooth operations and functionality of the energy systems within the premises or remotely. The difficulties occur during data transfer between these devices; for example, FDI might target the falsification of data transmitted between smart grids. It would hijack the session or communication channel with malicious packets and disable sensor nodes to damage networks and other hardware systems [14].

To minimize malicious data tampering and destruction, smart grids necessitate that data be of high integrity. A common integrity attack scenario, according to Cui et al [11], would involve changing settings in a power plant and broadcasting a data packet to the state estimator. In this case, the security posture is to have a non-repudiation and authentication mechanism in place to ensure that no party denies participating in an operation and that everyone using the system must authenticate her identity.

Conversely, if the data transmission network is about to push the system over the brink, fraudulent data injection might certainly occur. Due to system failure or unavailability, data may be forced to be offline, making any offline malicious packet injection by a physical intruder or disgruntled employee on the premises difficult to identify. More crucially, in smart grids or any other system, data confidentiality governs disclosure. If all data and communication channels are adequately secured with the appropriate cryptographic techniques, fake data injection would be difficult to achieve. According to Lee et al. [15], if the smart grids' confidentiality is revealed, sensitive data such as sensor control parameters and billing credentials would be compromised, causing financial and reputational damage to users and service providers.

Several machine-learning techniques to non-technical losses, such as energy theft detection, have been presented, with the supervised approach with SVM being preferred over other algorithms [16]. In those studies, the SVM was integrated with other algorithms for voltage sensitivity analysis and power optimization. Other methods, such as K-nearest neighbors (KNN), regression, and artificial neural networks (ANN), have also been explored and shown to produce similar results [17, 18, 19]. However, supervised approaches appear to produce a high number of false positives, which has a significant impact on the detection rate. Furthermore, categorizing the datasets for this purpose was a difficult undertaking. As a result, other researchers concentrated on the unsupervised learning strategy as an alternative. Clustering-based algorithms were the most prevalent method.

However, the unsupervised approach of clustering has a low detection rate with tampered data, thus Zheng et al. [20] suggested a detection framework that integrates state-based methods with clustering. Other studies used gradient boosting (extreme gradient boosting, category boosting, and light gradient boosting) to investigate the aforementioned approach [22, 23, 24].

The majority of the machine-learning algorithms used in smart grids are still hampered by a number of issues. High false positive rates due to underfitting and class imbalance in training datasets are just a few of the issues. As a result, several attempts have been made to detect fake-normal traffic using deep-learning algorithms, and one possible research topic would be to efficiently detect fraudulent data injection. Convolutional neural networks (CNN) have been explored extensively, among other things. For example, [25, 26, 27] used deep-learning methodologies to construct the detection algorithm with varied datasets.

IV. POTENTIAL FUTURE RESEARCH

Although a variety of techniques have been used to secure smart grid systems, only a few have focused on safeguarding smart grid systems from assaults designed to deceive AI-based models. Most existing research, for example, focuses on improving the performance of machine/deep learning models to achieve higher detection accuracy while maintaining reasonable computational performance, but it pays little attention to the fake-normal data traffic generated by Generative Adversarial Networks (GAN). They also don't spend much time examining the subtle variations between fake-normal and actual traffic.

By combining a generator and a discriminator, GANs are particularly effective in fooling artificial intelligence systems. The attacker tries to mimic the prediction function that the smart grid's pre-trained ML model uses to detect attacks by employing the discriminator. Also, utilize the generator to create benign traffic from malicious traffic. This is accomplished by a series of iterations in which the generator or (GAN) adds noise and fine-tunes its malicious packets/traffic to the point where they elude the smart grid model's detection radar. As a result, the attacker's traffic is able to pass through the algorithm without being detected. Instead of being classified as malicious, they would be classified as normal traffic.

Finally, I'd like to suggest that GAN attacks on smart grid systems be investigated further in order to make detectors more resilient against intelligent attacks.